\documentclass[secnumarabi,superscriptaddress,amssymb,nobibnotes,aps,prc]{revtex4}

\newcommand{\Pomeron}{I\!\!P}

\usepackage{epsfig}
\usepackage{graphicx}
\usepackage{amsmath}
\usepackage{hyperref}

\begin{document} 

\title{Accessing transverse nucleon and gluon distributions
in heavy nuclei using coherent vector meson photoproduction at high energies in ion ultraperipheral collisions}

\author{V. Guzey}
\affiliation{National Research Center ``Kurchatov Institute'', Petersburg Nuclear Physics Institute (PNPI), Gatchina, 188300, Russia}

\author{M. Strikman}
\affiliation{Department of Physics, The Pennsylvania State University, State College, PA 16802, USA}

\author{M. Zhalov}
\affiliation{National Research Center ``Kurchatov Institute'', Petersburg Nuclear Physics Institute (PNPI), Gatchina, 188300, Russia}


\begin{abstract} 

By using the theoretical approaches describing well the available data on $t$-integrated 
coherent photoproduction of light and heavy vector mesons in Pb-Pb ultraperipheral collisions (UPCs) at the Large Hadron Collider (LHC) in Run 1, 
we calculate the momentum transfer distributions for this process for $\rho$ and $J/\psi$ vector mesons in the 
kinematics of Run 2 at the LHC. 
We demonstrate that nuclear shadowing not only suppresses the absolute value of the cross sections, 
but also shifts the momentum transfer distributions toward smaller values of the momentum transfer $|t|$. 
This result can be interpreted as a broadening in the impact parameter space of the effective nucleon density in nuclei by 14\% in the case of 
$\rho$ and the nuclear gluon distribution by $5-11$\% in the case of $J/\psi$. 

\end{abstract}

\maketitle 

\section{Introduction}
\label{sec:Intro}

High-energy exclusive (elastic) processes with various beams and targets provide information on the distribution of scattering centers in the target in the plane perpendicular to the beam direction 
(the impact parameter plane), which makes possible the transverse imaging of the target. Its examples are well-known and numerous. 
Measuring the intermediate energy elastic scattering of 
electrons and protons on nuclei allows one to reconstruct the charge and matter 
(proton+neutron) distributions in nuclei, respectively.
The data on elastic proton--proton scattering at high energies are widely used to learn about
the proton profile in the impact parameter space.
The recent relativistic analyses of elastic form factors of 
hadrons (proton, neutron, pion) measured in elastic scattering
were carried out in terms of the transverse quark densities.
More generally, generalized parton distribution functions (GPDs) accessed in hard exclusive processes encode 
information of the quark and gluon distributions in a given target (including transitions between the 
hadronic states) in terms of light-cone momentum fractions and the impact parameter. Thus, they at least in principle 
hold the promise for obtaining a three-dimensional picture of hadrons and nucleus in Quantum Chromodynamics (QCD), 
which is one of the key objectives of the physics program of 
 a future Electron-Ion Collider~\cite{Accardi:2012qut}.

Part of this program involving high energy quasireal photons
overlaps with studies of proton--proton, proton--nucleus and nucleus--nucleus ultraperipheral collisions (UPCs)~\cite{Baltz:2007kq}
at the Large Hadron Collider (LHC).
In particular, during Run 1 in Pb-Pb UPCs, the ALICE collaboration~\cite{Adam:2015gsa} measured coherent photoproduction of $\rho$ 
mesons on heavy nuclei and the ALICE~\cite{Abbas:2013oua,Abelev:2012ba}
and CMS~\cite{Khachatryan:2016qhq} collaborations measured coherent $J/\psi$ photoproduction on nuclei.
Comparison of these data with numerous model calculations~\cite{Rebyakova:2011vf,Frankfurt:2015cwa,Santos:2014vwa,Adeluyi:2012ph,Guzey:2013xba,Guzey:2013qza,Lappi:2013am,Goncalves:2014wna} demonstrated 
that the best agreement is observed only if the calculations account for the effect of strong nuclear shadowing,
which suppresses the $\rho$ photoproduction cross section by approximately a factor of six and the
$J/\psi$ cross section by approximately a factor of three.
In particular, it was shown that the Gribov--Glauber approach to nuclear shadowing combined with phenomenology of 
real and virtual photon diffraction~\cite{Frankfurt:2015cwa,Guzey:2013xba} provides a good description of the data.
While the statistics of $\rho$ and $J/\psi$ photoproduction in Run I was insufficient for the detailed study 
of the transverse momentum distributions of vector mesons in these processes, some hints of such shifting
have been observed by both ALICE~\cite{Adam:2015gsa} and STAR~\cite{Debbe:2012aa,Klein:2016dtn}
collaborations.  
Note that the preliminary PHENIX data on coherent and incoherent $J/\psi$ photoproduction in Au-Au UPCs at
$\sqrt{s_{NN}}=200$ GeV accompanied by forward neutron emission~\cite{Takahara:2012pza}
probes the nuclear gluon distribution at the momentum fraction $x \approx 0.015$, where the nuclear
shadowing suppression is not large; see the good description of this data in Ref.~\cite{Guzey:2013jaa}.
At the same time, the limited precision of the data and its wide binning in $t$ along with the expected small gluon nuclear shadowing make it very difficult to study the modification of the $t$ distribution discussed in this work. 
The much higher statistics of Run 2 at the LHC 
allows one to study the influence of strong nuclear shadowing on the transverse momentum distributions.

In this paper, by using the theoretical approaches describing well the available data from Run 1, 
we extend our previous study~\cite{Guzey:2016piu} and calculate the momentum transfer distributions for 
coherent photoproduction of $\rho$ and $J/\psi$ vector mesons on nuclei in Pb-Pb UPCs in Run 2 at the LHC.
We show that nuclear shadowing not only suppresses the absolute value of the cross sections, 
but also changes the shape of the differential cross sections by shifting 
the momentum transfer distributions toward smaller values of the momentum transfer $|t|$. 
 One can  interpret these results as an effective broadening in the impact 
parameter space of the nucleon density in nuclei in the case of  $\rho$ and the nuclear gluon distribution 
in the case of $J/\psi$. It is a generic and model-independent consequence of the fact that nuclear shadowing 
suppression at small impact parameters is stronger than that at the nucleus periphery.

\section{Coherent photoproduction of light and heavy vector mesons on nuclei}
\label{sec:rho}

Collisions of ions at large impact parameters, which are called ultraperipheral collisions (UPCs) in the literature,
give an opportunity to study photon-initiated processes at unprecedentedly high energies~\cite{Baltz:2007kq}.
The cross section of coherent vector meson $V$ photoproduction in nucleus--nucleus UPCs reads:
\begin{equation}
\frac{d\sigma_{AA \to V AA}(y)}{dy dt}=
N_{\gamma/A}(y) \frac{d \sigma_{\gamma A \to V A}(y)}{dt}+
N_{\gamma/A}(-y) \frac{d \sigma_{\gamma A \to V A}(-y)}{dt} \,,
\label{eq:cs}
\end{equation}
where $N_{\gamma/A}$ is the flux of equivalent photons emitted by either 
of the nuclei; $d \sigma_{\gamma A \to V A}(y)/dt$ 
is the differential cross section of coherent vector meson 
photoproduction on nuclei; $y$ is the vector meson 
rapidity and $t$ is the invariant momentum transfer squared. 
For a given $y$, the photon has the energy of 
$\omega=(M_V/2) e^{y}$, when emitted by the nucleus moving in the 
direction of $V$, or the energy of  $\omega=(M_V/2) e^{-y}$, 
when emitted by the nucleus moving in the opposite direction.
Note that allowing the final nucleus to disintegrate in 
the $\gamma A \to V A^{\prime}$ process, one can 
use Eq.~(\ref{eq:cs})
to calculate also incoherent vector meson photoproduction on nuclei in UPCs.

In the $\rho$ meson case,
the coherent $\gamma A \to \rho A$ cross section in the approach based 
on the combination of the vector meson dominance 
(VMD) model and the Glauber model of nuclear shadowing 
(we collectively called it VMD-GM) reads~\cite{Bauer:1977iq}:
\begin{equation}
\frac{d \sigma_{\gamma A \to \rho A}^{\rm VMD-GM}(W_{\gamma p})}{dt} =
\left(\frac{e}{f_{\rho}}\right)^2 \frac{(1+\eta^2) \sigma_{\rho N}^2}{16 \pi}
\left|\int d^2 \vec{b} \, e^{i \vec{q}_{\perp} \vec{b}}  
\int dz \rho_A(b,z) e^{i q_{\parallel} z} e^{-\frac{1}{2}(1-i \eta) \sigma_{\rho N} \int_z^{\infty} dz^{\prime} 
\rho_A(b,z^{\prime})}\right|^2
\,.
\label{eq:vmd-gm}
\end{equation}
In Eq.~(\ref{eq:vmd-gm}),
$W_{\gamma p}$ is the invariant photon--nucleus energy per nucleon;
$\sigma_{\rho N}$ is the total vector meson--nucleon cross section;
$f_{\rho}$ is the  $\gamma-\rho$ coupling constant; 
$\eta$ is
the ratio of the real to the imaginary parts of the 
$\gamma p \to \rho p$ amplitude;
$\rho_A(b,z)$ is the nucleon density, which depends on the longitudinal ($z$) and transverse ($\vec{b}$) 
coordinates of the active nucleon in a nucleus; 
$\vec{q}_{\perp}$ 
and $q_{\parallel}$ are the transverse and longitudinal 
components of the momentum transfer, $t=-q_{\parallel}^2-\vec{q}_{\perp}^2$.
This approach and its generalizations~\cite{Frankfurt:2002wc,Frankfurt:2002sv}
taking into account the
influence of the higher $\rho^\prime$ component of the photon
wave function and nondiagonal $\rho\leftrightarrow \rho^\prime$ transitions
 provide a good description 
of the available fixed-target data and the RHIC 
data on $\rho$ photoproduction in Au-Au UPCs at $\sqrt{s_{NN}}=62.4$ GeV and 130 GeV corresponding to $W_{\gamma p} \leq 10$ GeV~\cite{Agakishiev:2011me,Adler:2002sc}.

With an increase of the photon energy, the inclusion of only lower hadronic components ($\rho^\prime$) of the photon wave function becomes insufficient and one needs to  take into account 
the effect of inelastic (Gribov) nuclear shadowing corresponding 
to the photon diffraction into large masses.
In Ref.~\cite{Frankfurt:2015cwa}, this was realized by using the Good--Walker formalism of cross section fluctuations~\cite{Good:1960ba,Blaettel:1993ah} by introducing the distribution $P_V(\sigma)$ giving the probability 
for the photon to interact with a nucleon target with the cross section $\sigma$. 
Note that in addition to significant inelastic nuclear shadowing, the cross section fluctuations result in the reduction of the effective $\rho$ 
meson--nucleon cross section probed in the $\gamma p \to \rho p$ process
compared to the additive quark model estimate. The resulting approach, which was called the modified VMD--Glauber--Gribov model
(mVMD-GGM) in  Ref.~\cite{Frankfurt:2015cwa},
leads to good description of all available data on coherent $\rho$ photoproduction on nuclei, including the 
RHIC Au-Au UPC data at $\sqrt{s_{NN}}=200$ GeV~\cite{Abelev:2007nb} and the LHC Pb-Pb UPC data at $\sqrt{s_{NN}}=2.76$ 
TeV~\cite{Adam:2015gsa} 
corresponding to $W_{\gamma p} > 10$ GeV.

Neglecting the longitudinal momentum transfer, 
the $\gamma A \to \rho A$ cross section in the mVMD-GGM approach has the following form 
in the high-energy limit:
\begin{equation}
\frac{d \sigma_{\gamma A \to \rho A}^{\rm mVMD-GGM} (W_{\gamma p})}{dt}=
\left(\frac{e}{f_{\rho}}\right)^2 \frac{1}{4 \pi}
\left|\int d^2 \vec{b} \, e^{i \vec{q}_{\perp} \vec{b}} \int d\sigma P_V(\sigma) \left(1-e^{-\frac{1}{2}(1-i \eta) \sigma T_A(b)}\right)\right|^2
\,,
\label{eq:mvmd-ggm}
\end{equation}
where $T_A(b)=\int dz \rho_A(b,z)$ is the density of nucleons 
in the transverse plane (the nuclear optical 
density) and $P_V(\sigma)$ is the distribution over cross section 
fluctuations for the $\gamma N \to \rho N$ transition.
This normalized distribution is centered around the effective $\rho$ 
meson--nucleon cross section extracted from the 
HERA $\gamma p \to \rho p$ data and has the dispersion determined 
by the photon diffractive dissociation $\gamma p \to Xp$
into large masses measured at the Fermilab, see details in Ref.~\cite{Frankfurt:2015cwa}.  

Coherent nuclear scattering at $t \neq 0$ is always accompanied by incoherent and inelastic processes, which contaminate the elastic 
signal. Hence, in practice one needs to separate the coherent and incoherent signals and examine the 
smearing of diffractive minima by the incoherent contribution. 
In UPCs, incoherent processes are characterized by inelastic final states in the forward direction and can be separated
from the coherent process by examining the $t$ dependence. 
To address these issues, we calculate the incoherent 
$\gamma A \to \rho A^{\prime}$ cross section ($A^{\prime} \neq A$ denotes products of nuclear disintegration)
 by using the following well-known expression~\cite{Bauer:1977iq}:
\begin{equation}
\frac{d \sigma_{\gamma A \to \rho A^{\prime}}^{\rm VMD-GM}(W_{\gamma p})}{dt}=\frac{d \sigma_{\gamma p \to \rho p}(W_{\gamma p})}
{dt} \int d^2 \vec{b} \, T_A(b) e^{-\sigma_{\rho N}^{\rm in} T_A(b)} \,,
\label{eq:incoh}
\end{equation}
where $\sigma_{\rho N}^{\rm in}=\sigma_{\rho N}-\sigma_{\rho N}^2/(16 \pi B_{\rho})$ is the inelastic $\rho$--nucleon cross section;
$B_{\rho}=  10.9 +0.46 \ln (W_{\gamma p}/72 \ {\rm GeV})^2$ GeV$^{-2}$ is the slope of the $t$ dependence of the $\gamma p \to \rho p$ cross section
at the relevant energies~\cite{Aid:1996bs,Breitweg:1997ed}.
Note that in Eq.~(\ref{eq:incoh}) we neglected the effect of cross section fluctuations and, thus, 
obtained the upper limit on the inelastic cross section. 
The key feature of incoherent photoproduction on nuclear targets is that the $t$ dependence of the 
nuclear cross section is dictated by the $t$ dependence of photoproduction on the nucleon:
\begin{equation}
{d \sigma_{\gamma p \to \rho p}(W_{\gamma p},t)\over {dt}}=\left(\frac{e}{f_{\rho}}\right)^2 \frac{(1+\eta^2) \sigma_{\rho N}^2}{16 \pi}e^{B_{\rho}t} \,.
\label{eq:incoh_elem}
\end{equation}

Turning to the $J/\psi$ case, we notice that 
the main interest in studying exclusive $J/\psi$ photoproduction at high energies is that it gives an almost 
direct access to the gluon distribution $g_{T}(x,\mu^2)$
in a given target~\cite{Ryskin:1992ui,Brodsky:1994kf} at the resolution scale of
$\mu^2={\cal O}(m_c^2)$ ($m_c$ is the mass of the
charm quark) in the kinematical region, where the gluons carry a small 
fraction $x=M_{J/\psi}^2/W_{\gamma p}^2 \ll 1$ of the target momentum. 
To the leading orders in the strong coupling constant and the non-relativistic expansion for the $J/\psi$ 
distribution amplitude,
the cross section of coherent $J/\psi$ photoproduction on nuclei in the high-energy limit 
is usually written in the following form, see, e.g., Ref.~\cite{Guzey:2013qza}:
\begin{equation}
\frac{d\sigma_{\gamma A \to J/\psi A}}{dt} = \frac{d\sigma_{\gamma p \to J/\psi p}(t=0)}{dt} 
\left(\frac{R_{g,A}}{R_{g,p}}\right)^2
\left(\frac{g_A(x,\mu^2)}{A g_p(x,\mu^2)} \right)^2 F_A^2(t)  \,,
\label{eq:Jpsi_2_app}
\end{equation}
where $d\sigma_{\gamma p \to J/\psi p}(t=0)/dt$ is the differential cross section on the proton at $t \approx 0$;
$g_A(x,\mu^2)/[A g_p(x,\mu^2)]$ is the ratio of the nuclear and proton gluon distributions;
$F_A(t)=\int d^2 b\, e^{i \vec{q}_{\perp} \vec{b}} T_A(b)$ is the nuclear form factor; 
$R_{g,A}$ and $R_{g,p}$ are the so-called skewness factors for the nucleus and proton gluon GPDs, respectively, 
which can be estimated by using the small-$x$ behavior of the corresponding gluon distributions~\cite{Shuvaev:1999ce}.
Note that while the leading order description of $J/\psi$ photoproduction is subject to sizable corrections~\cite{Ryskin:1995hz,Frankfurt:1997fj},
we expect that they largely cancel in Eq.~(\ref{eq:Jpsi_2_app}).

Equation~(\ref{eq:Jpsi_2_app}) assumes that nuclear shadowing does not affect the $t$ dependence of 
$d\sigma_{\gamma A \to J/\psi A}/dt$, which is then given by the nuclear form factor squared $F_A^2(t)$.
This is an approximation valid, when the effect of nuclear shadowing is insignificant. However, as we discussed in the 
Introduction, this is not the case. Hence, strong nuclear shadowing should
also modify the $t$ dependence of the  $d\sigma_{\gamma A \to J/\psi A}/dt$ cross section. Generalizing 
Eq.~(\ref{eq:Jpsi_2_app}) to the case of large nuclear shadowing, we obtain:
\begin{equation}
\frac{d\sigma_{\gamma A \to J/\psi A}}{dt}=\frac{d\sigma_{\gamma p \to J/\psi p}(t=0)}{dt} \left(\frac{R_{g,A}}{R_{g,p}}\right)^2
\left(\frac{g_A(x,t,\mu^2)}{Ag_p(x,\mu^2)}\right)^2 \,,
\label{eq:Jpsi}
\end{equation}
where $g_A(x,t,\mu^2)$ is the nucleus generalized gluon distribution in the special limit, when both gluon lines carry 
the equal light-cone momentum fractions of $x$.
In this case, GPDs can be expressed in terms of the impact parameter dependent PDFs~\cite{Burkardt:2000za}. 
In particular, we have for $g_A(x,t,\mu^2)$:
\begin{equation}
g_A(x,t,\mu^2)=\int d^2 \vec{b} \, e^{i \vec{q}_{\perp} \vec{b}} g_A(x,b,\mu^2) \,,
\label{eq:Jpsi_ft}
\end{equation}
where $xg_A(x,b,\mu^2)$ is the impact parameter dependent gluon nuclear PDF~\cite{Frankfurt:2011cs,Helenius:2012wd}, which
gives the probability to find in a nucleus a gluon with the light-cone momentum fraction $x$ 
at the transverse distance $b$ from the nucleus (center-of-mass) center. 
Using the results of Ref.~\cite{Guzey:2013jaa} for the calculation of $g_A(x,b,\mu^2)$, the 
expression for the cross section of coherent $J/\psi$ photoproduction on nuclei reads:
\begin{equation}
\frac{d\sigma_{\gamma A \to J/\psi A}}{dt}=\frac{d\sigma_{\gamma p \to J/\psi p}(t=0)}{dt} 
\left|\int d^2 \vec{b} \, e^{i \vec{q}_{\perp} \vec{b}}\left[\left(1-\frac{\sigma_2}{\sigma_3}\right) T_A(b)+\frac{2 \sigma_2}{\sigma_3^2} \Re e
\left(1-e^{-\frac{1}{2} \sigma_3(1-i\eta) T_A(b)}\right)\right] \right|^2 \,.
\label{eq:Jpsi_2}
\end{equation}
Equation~(\ref{eq:Jpsi_2}) is a series in powers of $T_A(b)$ (numbers
of interactions with target nucleons), which builds the  nuclear shadowing suppression.
The term proportional to $T_A^2(b)$ describes the interaction with two nucleons, whose strength is given by the $\sigma_2$ cross
section:
\begin{equation}
\sigma_2=\frac{16 \pi B_{\rm diff}}{(1+\eta^2)  xg_p(x,\mu^2)} \int_x^{0.1} dx_{\Pomeron} 
\beta g_p^{D(3)}(\beta,x_{\Pomeron},\mu^2) \,,
\label{eq:sigma_2}
\end{equation}
where $B_{\rm diff} \approx 6$ GeV$^{-2}$ is the slope of the $t$ dependence of the cross section of hard inclusive
diffraction on the proton in deep inelastic scattering (DIS) 
$\gamma^{\ast} p \to X p$~\cite{Aktas:2006hx}; 
$\eta \approx 0.17$ is the ratio of the real to imaginary parts of the $\gamma^{\ast} p \to X p$ amplitude estimated 
by using
the Gribov--Migdal relation;
$g_p^{D(3)}(\beta,x_{\Pomeron},\mu^2)$ is the gluon diffractive parton distribution of the 
proton~\cite{Aktas:2006hx,Aktas:2006hy}, which depends on the diffractive exchange (``Pomeron'') momentum fraction
$x_{\Pomeron}$, the gluon momentum fraction $\beta$, and the scale $\mu^2$.
By using the leading order (LO) H1 diffractive PDFs and the CTEQ6L1 gluon density~\cite{Pumplin:2002vw}, we find 
that $\sigma_2 = 21$ mb at $x=0.001$ and $\mu^2=3$ GeV$^2$.
The interaction with three and more nucleons is modeled by the effective cross section 
of $\sigma_3$, which is constrained by using the formalism of cross section fluctuations in Ref.~\cite{Frankfurt:2011cs}. We estimate that $\sigma_3=26-45$ mb, 
which presents the main source of theoretical uncertainties on $g_A(x,b,\mu^2)$ in this approach.
One can see from Eq.~(\ref{eq:Jpsi_2}) that nuclear shadowing (the sum of terms proportional to $T_A^n(b)$, with $n \geq 2$) affects not only the magnitude of $\sigma_{\gamma A \to J/\psi A}$, but also
the $t$ dependence of $d\sigma_{\gamma A \to J/\psi A}/dt$. 

It is important to note that while Eqs.~(\ref{eq:Jpsi_2_app}) and (\ref{eq:Jpsi_2}) give discernibly different predictions 
for $d\sigma_{\gamma A \to J/\psi A}/dt$, their predictions for the $t$ integrated $\sigma_{\gamma A \to J/\psi A}$
cross section differ by less than approximately 15\% (the approximate expression 
gives the larger cross section than the exact one).
This gives a less than 8\% correction to the nuclear suppression factor of $S_{Pb}$ 
for coherent $J/\psi$ photoproduction in Pb-Pb UPCs
predicted using~Eq.~(\ref{eq:Jpsi_2_app})~\cite{Guzey:2013xba,Guzey:2013qza}, which is comparable to 
the experimental uncertainty of 
the LHC data for this process~\cite{Abbas:2013oua,Abelev:2012ba,Khachatryan:2016qhq}.

With the same formalism, we estimate the cross section of incoherent $J/\psi$ photoproduction on nuclear 
targets~\cite{Guzey:2013jaa}:
\begin{equation}  
\frac{\sigma_{\gamma A\rightarrow J/\psi A^{\prime}}^{\rm pQCD}(W_{\gamma p})}{dt}=
\frac{d \sigma_{\gamma p \to J/\psi p}(W_{\gamma p})}
{dt} \int d^2 \vec{b} \, T_A(b) \left [1-\frac {\sigma_2} {\sigma_3} \left (1- 
e^{-{\frac {\sigma_3} {2}} T_A(b)}\right )\right ]^2 \,.
\label{eq:sigma_incoh_QCD}
\end{equation}
For the $t$ dependence of the elementary $\gamma p \to J/\psi p$ cross section, we use the following simple
exponential form:
\begin{equation}  
\frac{d \sigma_{\gamma p \to J/\psi p}(W_{\gamma p})}{dt}=
\frac{d\sigma_{\gamma p \to J/\psi p}(t=0)}{dt} e^{B_{J/\psi}t} \,,
\label{eq:JPsi_p_tdep}
\end{equation}
where $B_{J/\psi}(W_{\gamma p})=4.5+0.4 \ln(W_{\gamma p}/90\ {\rm GeV})$, which describes well 
the HERA data on the $t$ dependence of the cross section of $J/\psi$ photoproduction on the proton, 
see, e.g.~\cite{Guzey:2013qza}.

\section{Results and discussion}
\label{sec:results}

\begin{figure}[t]
\begin{center}
\epsfig{file=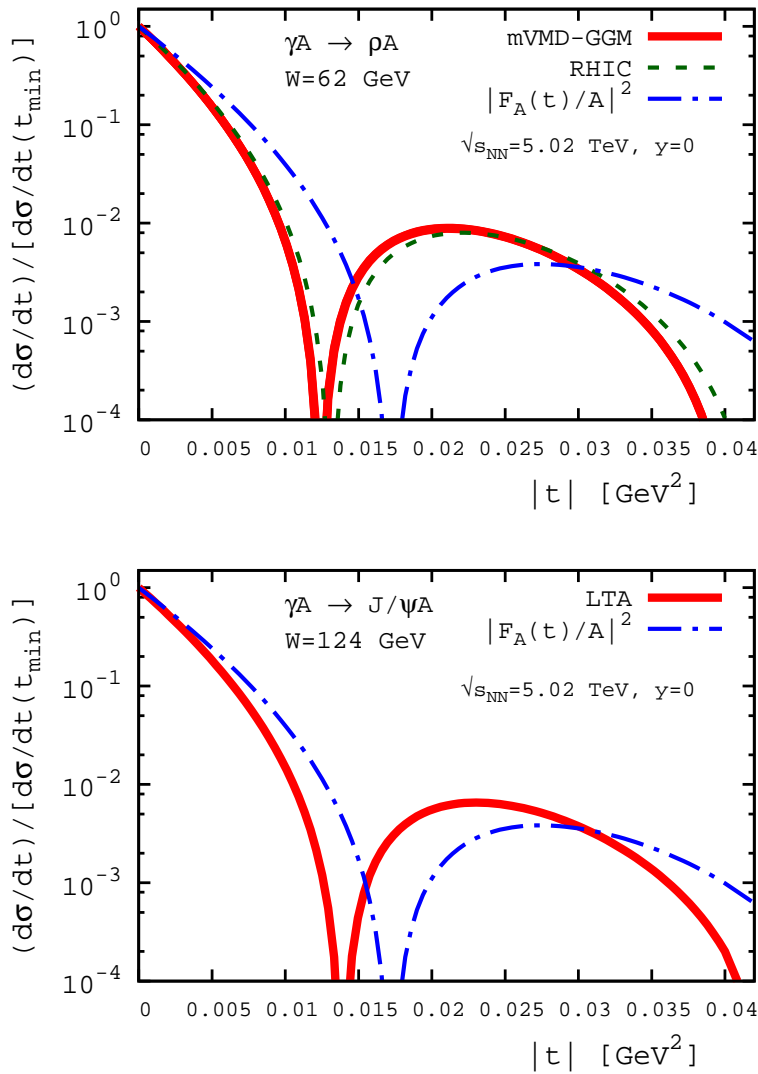,scale=1.4}
\caption{
The $d \sigma_{\gamma A \to V A}(W_{\gamma p})/dt$ cross section for $\rho$ (top panel) and $J/\psi$ (lower panel) for $^{208}$Pb normalized to its value at $t=t_{\rm min}$ as a function of $|t|$.
The cross section are calculated at $W_{\gamma p}=62$ GeV for $\rho$ and $W_{\gamma p}=124$ GeV for $J/\psi$, 
corresponding to the LHC Run 2 $\sqrt{s_{NN}}=5.02$ TeV and $y=0$.
The resulting $t$ dependence is compared to that given by the normalized nuclear form factor squared $|F_A(t)/A|^2$.
For $\rho$ meson, we also show the result of the calculation at $W_{\gamma p}=10$ GeV corresponding to the RHIC
kinematics (the green dashed line labeled ``RHIC'').
}
\label{fig:rho_t_new}
\end{center}
\end{figure}

Figure~\ref{fig:rho_t_new} shows our results for the 
$d \sigma_{\gamma A \to V A}(W_{\gamma p})/dt$ cross section for $\rho$ (top panel) and $J/\psi$ (lower panel)
coherent photoproduction on $^{208}$Pb as a function of $|t|$. The cross sections are normalized to their values at $t=t_{\rm min}$, where $t_{\rm min}=-m_N^2 M_{\rho}^4/W^4_{\gamma p}$,
and are evaluated at $W_{\gamma p}=62$ GeV for $\rho$ and $W_{\gamma p}=124$ GeV for $J/\psi$, which
corresponds to $y=0$ for Pb-Pb UPCs at $\sqrt{s_{NN}}=5.02$ TeV.
In the upper panel, the red solid curve labeled ``mVMD-GGM'' corresponds to Eq.~(\ref{eq:mvmd-ggm}).
In the bottom panel, the red solid curve labeled ``LTA'' shows the result of Eq.~(\ref{eq:Jpsi_2}) calculated 
with the lower value of $\sigma_3$, which corresponds to the upper limit on the shadowing effect for $J/\psi$ photoproduction.
For reference, we also show the normalized nuclear form factor squared obtained by using the nucleon
density of $^{208}$Pb of Ref.~\cite{DeJager:1987qc} (the blue dot-dashed curve labeled ``$|F_A(t)/A|^2$'').
In the $\rho$ meson case, we also show the result of the calculation at $W_{\gamma p}=10$ GeV corresponding to the RHIC
kinematics (the green dashed line labeled ``RHIC''). One can see that the normalized momentum transfer distribution
is a weak function of $W_{\gamma p}$ between the RHIC and LHC energies.

One can see from the figure that nuclear shadowing modifies the $t$ dependence of $d \sigma_{\gamma A \to V A}(W_{\gamma p})/dt$
 by shifting the positions of the diffractive minima
 and maxima towards smaller values of $|t|$. 
 For instance, the shift of the first minimum is $\Delta p_t \approx 18$ MeV for $\rho$ and $\Delta p_t \approx 14$ MeV for 
 $J/\psi$. 
 Note that in the $\rho$ meson case, the predicted $t$ dependence very weakly depends on details of the model of
 cross section fluctuations.
 In the $J/\psi$ case, the effect of cross section fluctuations is implicit in Eq.~(\ref{eq:Jpsi_2})
 and the $\Delta p_t$ shift depends on the value of the average $\sigma_3$ cross section, which 
 has a significant uncertainty and
  constrained to lie
 in the $\sigma_3=26-45$ mb interval.
 The result of the calculation with the lower value of $\sigma_3$, which corresponds to the scenario with the 
 larger gluon shadowing in the leading
 twist model of nuclear shadowing~\cite{Frankfurt:2011cs}, is presented in Fig.~\ref{fig:rho_t_new}.
 For the larger value of $\sigma_3$ and the correspondingly smaller gluon shadowing, the modification
 of the $t$ distribution of $d \sigma_{\gamma A \to J/\psi A}(W_{\gamma p})/dt$ compared to $|F_A(t)/A|^2$ is smaller;
 the corresponding shift is $\Delta p_t \approx 6$ MeV. 

The shift of the $t$ dependence of the $d \sigma_{\gamma A \to V A}(W_{\gamma p})/dt$ cross section shown in 
Fig.~\ref{fig:rho_t_new} can be interpreted as an increase (broadening) in the 
impact parameter space of the nucleon density in nuclei in the case of  $\rho$ and the nuclear gluon distribution 
in the case of $J/\psi$.
Characterizing the average transverse size of these distributions 
by the equivalent radius of $R_A$, one can estimate the relative increase of $R_A$ as $\Delta R_A/R_A \approx \Delta p_t/p_t$, 
which gives $\Delta R_A/R_A \approx 1.14$ for $\rho$ and $\Delta R_A/R_A \approx 1.05-1.11$ for $J/\psi$.
The latter estimate agrees with the results of the analysis of the average transverse size of the 
nuclear gluon distribution of Ref.~\cite{Frankfurt:2011cs}.
The transverse broadening of the nuclear gluon and sea quark distributions caused by nuclear shadowing can also be studied 
in other exclusive processes such as, e.g., deeply virtual Compton scattering, where it leads to dramatic oscillations
of the beam-spin cross section asymmetry~\cite{Frankfurt:2011cs}.

\begin{figure}[t]
\begin{center}
\epsfig{file=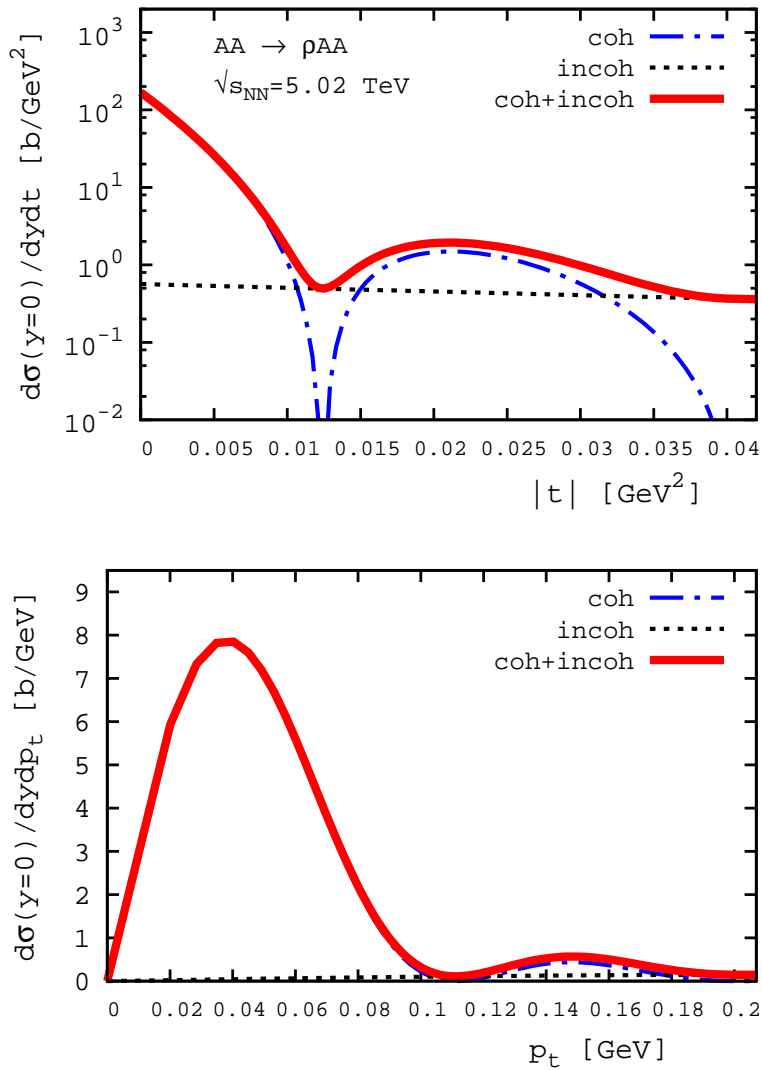,scale=1.4}
 \caption{Photoproduction of $\rho$ mesons in Pb-Pb UPCs at $y=0$ and  $\sqrt{s_{NN}}=5.02$ TeV:  
 $d\sigma_{AA \to \rho AA}(y=0)/dy dt$ as a function of 
$|t|$ (top panel) and $d\sigma_{AA \to \rho AA}(y=0)/dy dp_t$ as a function of 
$p_t$ (bottom panel). 
The blue dot-dashed and black dotted curves give separately the coherent and incoherent contributions, while the red solid curve 
is their sum.
}
\label{fig:cscindpt}
\end{center}
\end{figure}

Figure~\ref{fig:cscindpt} shows our predictions for $d\sigma_{AA \to \rho A^{\prime}A}(y=0)/dy dt$ as a function of 
$|t|$ (top panel) and $d\sigma_{AA \to \rho A^{\prime}A}(y=0)/dy dp_t$ as a function of 
$p_t$ (bottom panel) for Pb-Pb UPCs at $\sqrt{s_{NN}}=5.02$ TeV for Run 2 at the LHC
($A^{\prime}$ denotes both coherent $A^{\prime}=A$ and incoherent $A^{\prime} \neq A$ cases).
The blue dot-dashed and black dotted curves give the coherent [Eqs.~(\ref{eq:cs}) and (\ref{eq:mvmd-ggm})]
and incoherent [Eqs.~(\ref{eq:incoh})] contributions, respectively; the red solid curve is the sum
of the coherent and incoherent terms. One can see from the figure that while the incoherent contribution partially fills in
the first diffractive minimum in the $t$ dependence, the minimum still remains visible and its position as a function of $|t|$ or 
$p_t$ is unaffected.

\begin{figure}[th]
\begin{center}
\epsfig{file=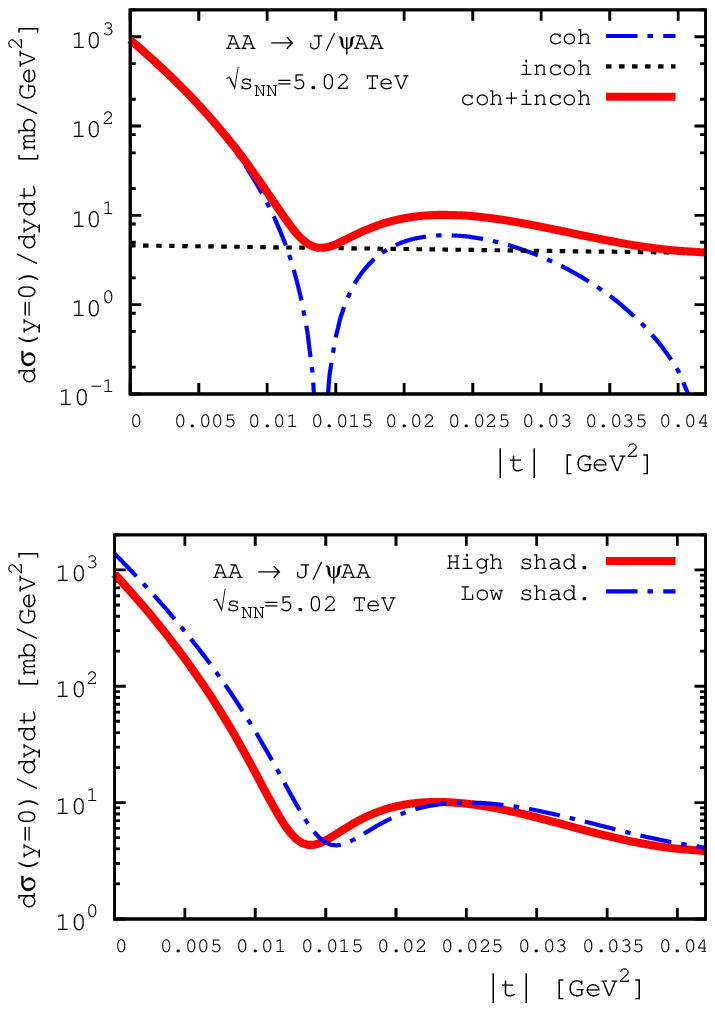,scale=1.4}
 \caption{
 Photoproduction of $J/\psi$ mesons in Pb-Pb UPCs at $y=0$ and $\sqrt{s_{NN}}=5.02$ TeV:
 $d\sigma_{AA \to J/\psi AA}(y=0)/dy dt$ as a function of $|t|$.
 Top panel: The blue dot-dashed and black dotted curves give separately the coherent and incoherent contributions
 calculated using the higher gluon shadowing,
  while the red solid curve is their sum. 
 Bottom panel: The sum of coherent and incoherent contributions calculated using the higher (red solid curve) and
lower (blue dot-dashed curve) gluon nuclear shadowing.
 }
 \label{fig:Jpsi_t_incoh}
\end{center}
\end{figure}

The differential $d\sigma_{AA \to J/\psi A^{\prime}A}(y=0)/dy dt$ cross section for $J/\psi$ photoproduction is shown in Fig.~\ref{fig:Jpsi_t_incoh}. The upper panel corresponds to the calculations with the higher 
leading twist gluon shadowing (smaller $\sigma_3$)~\cite{Frankfurt:2011cs} (as in Fig.~\ref{fig:rho_t_new}):
The blue dot-dashed and black dotted curves give separately the coherent and incoherent contributions, 
while the red solid curve is their sum.
In the lower panel, we compare the sum of coherent and incoherent contributions calculated by using the higher (the red solid curve) and lower (the blue dot-dashed curve) gluon nuclear shadowing.
One can see from the lower panel of the figure that the higher gluon shadowing leads to a larger shift of the 
$t$ distribution.
Also,  as in the $\rho$ meson case,
the incoherent contribution partially fills in the first diffractive minimum, which still remains visible.

Note that our results for the incoherent contribution to the $AA \to V A^{\prime}A$ cross section were derived using 
completeness of final nuclear states and neglecting inelastic $\gamma N \to V X$ processes on the nucleon~\cite{Guzey:2013jaa}.
This approach underestimates the measured $t$-integrated cross section of incoherent $J/\psi$ photoproduction
in Pb-Pb UPCs at the LHC at $\sqrt{s_{NN}}=2.76$ TeV~\cite{Abbas:2013oua} by approximately a factor of 1.5. 
Hence, a more accurate treatment of the incoherent contribution will somewhat increase its magnitude at the values of $t$
shown in Fig.~\ref{fig:Jpsi_t_incoh}, which will result in a less pronounced first diffractive minimum.

The standard method of separating the coherent and incoherent contributions is an examination of their $t$ ($p_t$) dependence;
this is illustrated in Figs.~\ref{fig:cscindpt} and \ref{fig:Jpsi_t_incoh}. In addition, one can experimentally 
suppress the incoherent contribution by using zero degree calorimeters registering forward neutrons. Since quasi-elastic scattering
leads to emission of one or more neutrons with 85\% probability, requiring that no neutrons are emitted (the so-called
0n0n-channel) suppresses the incoherent contribution at the 15\% level, see the discussion in Ref.~\cite{Guzey:2013jaa}.

\begin{figure}[t]
\begin{center}
\epsfig{file=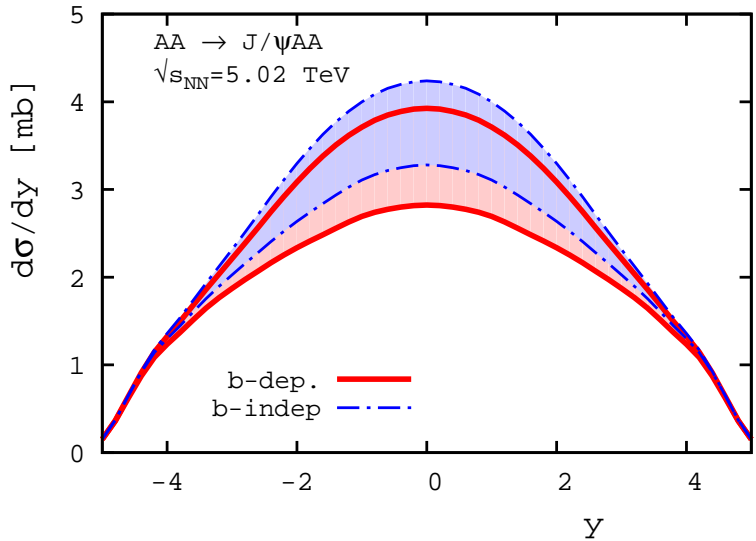,scale=1.4}
 \caption{
 The $J/\psi$ rapidity distribution in Pb-Pb UPCs at $\sqrt{s_{NN}}=5.02$ TeV:
 the lower band labeled `b-dep.'' corresponds to impact parameter dependent nuclear shadowing 
 [Eq.~(\ref{eq:Jpsi_2})], while the upper band labeled ``b-indep.'' is obtained using Eq.~(\ref{eq:Jpsi_2_app}) neglecting the non-trivial impact parameter dependence of gluon nuclear shadowing.
 }
 \label{fig:Sigma_PbPb_Run2}
\end{center}
\end{figure}

To further illustrate the effect of impact parameter dependent nuclear shadowing on the cross section of coherent $J/\psi$
photoproduction on nuclei, in Fig.~\ref{fig:Sigma_PbPb_Run2} we show our results for the $J/\psi$ rapidity distribution in 
Pb-Pb UPCs at $\sqrt{s_{NN}}=5.02$ TeV: the lower band labeled ``b-dep.'' is calculated using Eq.~(\ref{eq:Jpsi_2}), while the upper band labeled ``b-indep.'' is calculated using Eq.~(\ref{eq:Jpsi_2_app}).
One can see from the figure that taking into account the non-trivial impact parameter dependence of $g_A(x,b,\mu^2)$
somewhat lowers our predictions for $d\sigma_{AA \to J/\psi AA}(y)/dy$. For instance,  at $y=0$, we predict
that $d\sigma(y=0)/dy=2.82-3.93$ mb for the calculation with the impact parameter dependent shadowing 
(the ``b-dep.'' curves) and $d\sigma(y=0)/dy=3.28-4.24$ mb for the ``b-indep.'' case. It
corresponds to a 15\% reduction of $d\sigma(y=0)/dy$ for the higher gluon shadowing scenario (the lower boundary of the shaded bands in Fig.~\ref{fig:Sigma_PbPb_Run2}) and 8\% reduction of $d\sigma(y=0)/dy$ in the lower gluon shadowing case
(the upper boundary of the bands in Fig.~\ref{fig:Sigma_PbPb_Run2}).
As we already mentioned in Sect.~\ref{sec:rho}, this effect does not affect the good agreement between our 
earlier predictions~\cite{Guzey:2013xba,Guzey:2013qza} and the Run 1 LHC data~\cite{Abbas:2013oua,Abelev:2012ba,Khachatryan:2016qhq}.

\begin{figure}[th]
\begin{center}
\epsfig{file=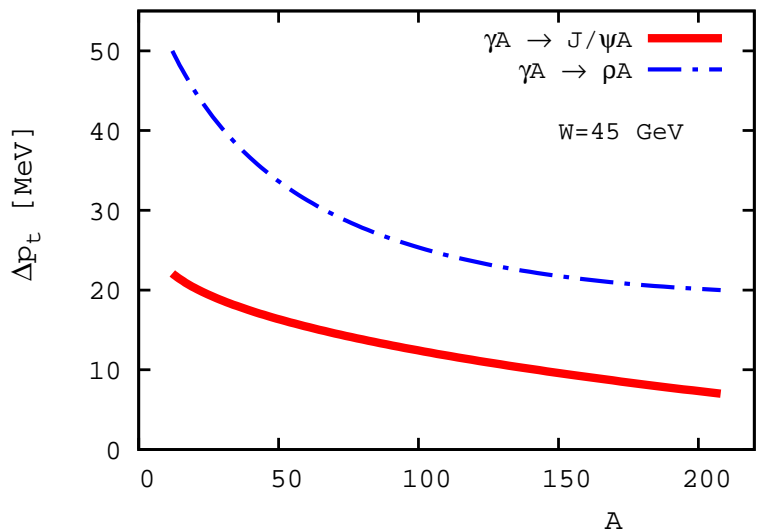,scale=1.4}
 \caption{The shift of the first diffractive minimum of $d\sigma_{\gamma A \to J/\psi A}/dt$ (red solid curve)
 and $d\sigma_{\gamma A \to \rho A}/dt$ (blue dot-dashed curve)
with the respect to that of $F_A^2(t)$, $\Delta p_t$, as a function of the atomic number $A$ at
$W_{\gamma p} = 45$ GeV. In the $J/\psi$ case, it corresponds to $x=M^2_{J/\psi}/W_{\gamma p}^2=0.005$.}
 \label{fig:pt_Adep}
\end{center}
\end{figure}

As we mentioned in the Introduction, UPC measurements at the LHC compliment the physics program of a future Electron-Ion 
Collider. To illustrate the EIC potential 
for transverse imaging
in real photon--nucleus scattering, 
in Fig.~\ref{fig:pt_Adep} we show the shift of the first diffractive minimum of the $d\sigma_{\gamma A \to J/\psi A}/dt$
(red solid curve) and $d\sigma_{\gamma A \to \rho A}/dt$ (blue dot-dashed curve)
cross sections with the respect to the first minimum of $F_A^2(t)$, $\Delta p_t$, as a function of the atomic number of $A$ at
$W_{\gamma p} = 45$ GeV. This value of $W_{\gamma p}$ conforms with the EIC kinematics and, in the case of $J/\psi$, 
corresponds to $x=M^2_{J/\psi}/W_{\gamma p}^2=0.005$.
One can see from the figure that the $\Delta p_t$ shift is sizable and increases with a decrease of $A$. 
The latter is a consequence of the fact that while the position of the first minimum of $F_A^2(t)$ scales as $A^{-1/3}$, 
the position of the first minimum of $d\sigma_{\gamma A \to V A}/dt$ scales somewhat slower due to nuclear 
shadowing, which makes $\Delta p_t$ a decreasing function of $A$.
Note that this $A$-behavior of $\Delta p_t$ changes for small $A \leq 4$, where one should approach the formal limit 
of $\Delta p_t \to 0$ for $A \to 1$. At the same time,  $\Delta p_t/p_t$ behaves monotonously and increases with an increase 
of $A$ for all atomic numbers.

In our analysis, we neglected the interference contribution in Eq.~(\ref{eq:cs}) and the resulting photon
contribution to the transverse momentum distributions~\cite{Klein:1999gv}. This contribution is confined to very low
$p_t < 10$ MeV and, hence, does not affect the results of our analysis focusing on the first diffractive minimum situated at much larger values of $p_t$.

Note that the shift of the diffractive minima due to nuclear shadowing that we observe is caused by soft diffraction in the case of $\rho$ and by leading twist hard diffraction in the case of $J/\psi$. 
In the $J/\psi$ case, this mechanism can be contrasted with predictions of other approaches available in the literature.
In the $k_t$-factorization approach~\cite{Cisek:2012yt}, the unintegrated nuclear gluon distribution receives 
contributions from multiple scattering of both quark--antiquark (so-called Glauber regime) and 
quark--antiquark--gluon dipoles on a nucleus, which determine the initial condition for the nonlinear evolution equation.
Note that the successful description of the proton diffractive structure functions measured in $ep$ DIS at HERA 
requires both quark-antiquark and quark-antiquark--gluon dipoles in the color dipole formalism;
it provides a connection to our leading twist approach, where the nuclear gluon shadowing is determined by the 
gluon diffractive parton distribution of the proton.
The inclusion of the quark--antiquark--gluon contribution leads to a noticeable suppression of the predicted 
impact parameter distribution of coherent $J/\psi$ photoproduction on Pb with an increase of the photon energy.
In the momentum space, it should correspond to a shift of the $t$ distribution toward smaller $|t|$, 
cf.~Ref.~\cite{Frankfurt:2011cs}.
Thus, regardless of the dynamical mechanism of nuclear shadowing, large nuclear gluon shadowing leads to the modification of 
the $t$ distribution of $J/\psi$ photoproduction in ion UPCs.
At the same time, in the implementations of the color dipole framework, where coherent photoproduction of $J/\psi$ 
on nuclei proceeds via multiple rescattering of quark--antiquark dipoles~\cite{Caldwell:2009ke,Toll:2012mb,Lappi:2013am}, 
the shadowing correction is not large since the average dipole--nucleon cross section is determined by the small
size of $J/\psi$. As a result, the modification of the $t$ distribution of $J/\psi$ photoproduction on nuclei compared to $F_A^2(t)$
 is smaller than predicted in our analysis.

\section{Conclusions}

In this paper, using the theoretical approaches describing well the available data on $t$-integrated 
coherent photoproduction of light and heavy vector mesons in Pb-Pb UPCs at the LHC during Run 1, 
we calculated the momentum transfer distributions for this process for $\rho$ and $J/\psi$ vector mesons 
in the kinematics of Run 2 at the LHC.
We demonstrated that nuclear shadowing not only suppresses the absolute value of the cross sections, 
but also shifts the momentum transfer distributions toward smaller values of the momentum transfer $|t|$. 
This result can be interpreted as a broadening in the impact parameter space of the effective nucleon density in nuclei in the case of 
$\rho$ and the nuclear gluon distribution in the case of $J/\psi$. 
Characterizing the average transverse size of these distributions 
by the equivalent radius of $R_A$, for the relative increase of $R_A$ we found 
$\Delta R_A/R_A \approx 1.14$ for $\rho$ and $\Delta R_A/R_A \approx 1.05-1.11$ for $J/\psi$.

The observed broadening of the transverse distributions is a model-independent consequence of nuclear shadowing, whose  
suppression effect at small impact parameters is stronger than at the nucleus periphery.
The transverse broadening of the nuclear gluon and sea quark distributions caused by nuclear shadowing can also be studied 
at EIC in such hard exclusive processes as, e.g., deeply virtual Compton scattering, where it leads to dramatic oscillations
of the beam-spin cross section asymmetry. All such measurements at the LHC and EIC will for the first time measure the 
impact parameter dependent quark and gluon distributions in nuclei and, hence, make an important step toward 
obtaining a three-dimensional image of parton distributions.

\acknowledgments

M.S.'s research was supported by the US Department of Energy Office of Science, Office of Nuclear Physics under Award 
No.~DE-FG02-93ER40771. V.G. would like to thank Pennsylvania State University for hospitality during the final stage of this work. 
M.Zh.'s research was supported in part by RF Ministry
of Science and Education under contract No. 14.610.21.0003.

\end{document}